\documentclass[12pt,preprint]{aastex}

\slugcomment{To appear in the 2005 Sep issue
             of the {\em Astronomical Journal.}}

\shorttitle{Faint Radio Sources in Bo\"otes} 
\shortauthors{Wrobel et al.}
\begin{document}

\title{Faint Radio Sources in the NOAO Bo\"otes Field.\\
       VLBA Imaging and Optical Identifications}

\author{J. M. Wrobel,\altaffilmark{1}
        G. B. Taylor,\altaffilmark{1,2}
        T. A. Rector,\altaffilmark{1,3}
        S. T. Myers,\altaffilmark{1}
    and C. D. Fassnacht\altaffilmark{4}}

\altaffiltext{1}{National Radio Astronomy Observatory, P.O. Box O,
Socorro, NM 87801; jwrobel@nrao.edu, gtaylor@nrao.edu, smyers@nrao.edu}

\altaffiltext{2}{Kavli Institute of Particle Astrophysics and
Cosmology, Menlo Park, CA 94025}

\altaffiltext{3}{Physics \& Astronomy Department, University of Alaska
Anchorage, 3211 Providence Drive, Anchorage, AK 99508;
aftar@uaa.alaska.edu}

\altaffiltext{4}{Department of Physics, University of California at
Davis, 1 Shields Avenue, Davis, CA 95616; cdfassnacht@ucdavis.edu}

\begin{abstract}
As a step toward investigating the parsec-scale properties of faint
extragalactic radio sources, the Very Long Baseline Array (VLBA) was
used at 5.0~GHz to obtain phase-referenced images of 76 sources in the
NOAO Bo\"otes field.  These 76 sources were selected from the FIRST
catalog to have peak flux densities above 10~mJy at 5\arcsec\,
resolution and deconvolved major diameters of less than 3\arcsec\, at
1.4~GHz.  Fifty-seven of these faint radio sources were identified
with accretion-powered radio galaxies and quasars brighter than
25.5~mag in the optical $I\/$ band.  On VLA scales at 1.4~GHz, a
measure of the compactness of the faint sources (the ratio of the peak
flux density from FIRST to the integrated flux density from the NVSS
catalog) spans the full range of possibilites arising from
source-resolution effects.  Thirty of the faint radio sources, or
39$^{+9}_{-7}$\%, were detected with the VLBA at 5.0~GHz with peak
flux densities above 6 $\sigma \sim$ 2~mJy at 2~mas resolution.  The
VLBA detections occur through the full range of compactness ratios.
The stronger VLBA detections can themselves serve as phase-reference
calibrators, boding well for opening up much of the radio sky to VLBA
imaging.  For the adopted cosmology, the VLBA resolution correponds to
17~pc or finer.  Most VLBA detections are unresolved or slightly
resolved but one is diffuse and five show either double or core-jet
structures; the properties of these latter six are discussed in
detail.  Three VLBA detections are unidentified and fainter than
25.5~mag in the optical $I\/$ band; their properties are highlighted
because they likely mark optically-obscured active nuclei at high
redshift.
\end{abstract}


\section{Introduction}

During the past decade, over two thousand radio continuum sources have
been surveyed at milliarcsecond (mas) resolution with the techniques
of very long baseline interferometry \citep[VLBI;
e.g.][]{bea02,fom03,pet05}.  Optical identifications and redshifts are
time-consuming to obtain and are available only for a few hundred
sources \citep{ver03,kel04}.  For those sources, the accepted paradigm
is that VLBI imaging traces the relativistic outflows on parsec scales
that emerge from the accretion regions of supermassive black holes.
These traditional surveys rely on self-calibration techniques and are
therefore biased by targeting sources brighter than about 50~mJy
\citep[e.g.][]{tay05}.  Moreover, these surveys only target sources
with flat radio spectra, to enhance the prospects that
synchroton-self-absorbed emission will be present on mas scales.  Do
the parsec-scale properties of the fainter radio population, selected
independent of radio spectra, differ from the properties inferred for
the brighter, flat-spectrum population?

Addressing this question is now feasible because of three recent
developments.  First, the technique of phase referencing is now in
routine use on VLBI arrays \citep{wro00},\footnote{Available at
http://www.vlba.nrao.edu/memos/sci/sci24memo.ps.} enabling astrometric
imaging of sources fainter than 50~mJy \citep[e.g.,][]{xu00}.  Second,
wide-area radio surveys with sub-arcminute resolution have defined
source populations at millijansky levels \citep[e.g.,][]{con99}.
Third, the prior diffiulties with obtaining optical identifications
and redshifts can be mitigated by focusing on faint radio sources
within known optical deep fields \citep[e.g.,][]{cri01}.  Combining
these three developments, the NRAO Very Long Baseline Array
\footnote{The VLBA is operated by the National Radio Astronomy
Observatory, a facility of the National Science Foundation, operated
under cooperative agreement by Associated Universities, Inc.}
\citep[VLBA;][]{nap94} was used to conduct phase-referenced
observations at 5.0~GHz of faint radio sources in the 9~deg$^2$ of the
NOAO Bo\"otes field \citep{jan99, jan05, dey05}.  The NOAO limit of $I
\sim 25.5$~(Vega)~mag means it is feasible that many of these faint
sources can be identified, dominantly with accretion-powered radio
galaxies and quasars \citep{jan99}, and efficient follow-up with a
multi-object fiber spectrograph is possible.  The idea for this type
of VLBI survey was first suggested by \citet{gar98}.  \citet{gio05}
are pursuing a different strategy that will, eventually, include some
sources too faint for VLBI self-calibration; that survey is targeting
a complete sample of 95 radio sources selected independently of the
presence of compact emission and at redshifts $z < 0.1$.  Also,
\citet{por04} have been using a single VLBI baseline
(Arecibo-Effelsberg) to perform a survey of approximately 1000 radio
sources, independent of source flux density, size or spectrum.

This paper is organized as follows.  Section~2 describes the sample
selection for the VLBA survey, including the identification status in
the $I\/$ band available from \citet{jan05}.  The VLBA observations
and calibration are described in Section~3, while Section~4 describes
the VLBA imaging strategies.  The discussion begins with an overview
of the VLBA survey (Section 5.1) then explores implications related to
the VLBA detection rate (Section 5.2), the VLBA parsec-scale
structures (Section 5.3), and the characteristics of the eight VLBA
detections unidentifed in the NOAO Bo\"otes survey (Section 5.4).
Section~6 closes with a summary of this paper and also outlines future
steps concerning this sample of faint radio sources (Rector et al.\
2005, in preparation).

To facilitate comparison with other analyses in the NOAO Bo\"otes
field \citep{eis04, hou05, hig05}, this paper will use Vega magnitudes
and a flat ${\rm \Lambda}$CDM cosmology with $H_0 =
71$~km~s$^{-1}$~Mpc$^{-1}$, $\Omega_m = 0.27$, and $\Omega_\Lambda =
0.73$ \citep{spe03}.  Also, all quoted uncertainties will correspond
to 1~$\sigma$ Gaussian errors, including those for Poisson counting
errors as tabulated by \citet{geh86}.

\section{Sample Selection}

The faint radio sources were selected from the 1999 July 21 version of
the FIRST catalog \citep[Faint Images of the Radio Sky at
Twenty-Centimeters;][]{whi97} to have peak flux densities above 10~mJy
at 5\arcsec\, resolution and deconvolved major diameters of less than
3\arcsec\, at 1.4~GHz.  Figure~1 shows the locations of these 76 faint
and compact sources, while Table~1 lists for each source the
position-encoded FIRST name in column~1 and the peak flux density from
FIRST at 1.4~GHz in column~2.  Column~3 gives the integrated flux
density at 1.4~GHz obtained from the 2002 July 18 version of the NVSS
catalog at 45\arcsec\, resolution \citep[NRAO VLA Sky
Survey;][]{con98}.  Each FIRST source has a localization ellipse
diameter at 5 $\sigma$ of about 1.0\arcsec, so it is prudent to ensure
that the VLBA search region covers a diameter of at least 1.0\arcsec.

A cross-correlation of the 76 FIRST sources with the NOAO $I\/$ band
catalog \citep{jan05} was made.  Only identifications within a total
offset of 3\arcsec\, were considered as potential matches, although
none were found with offsets between 2-3\arcsec.  Rector et al.\
(2005, in preparation) will discuss these NOAO identifications in
detail.  For now, we note only that 55 of these faint radio sources
have been identified with accretion-powered radio galaxies and
quasars.  Column~4 of Table~1 indicates the identification status in
the $I\/$ band of the FIRST sources.

On VLA scales at 1.4~GHz, the ratio of the peak flux density from
FIRST (column 2, Table~1) to the integrated flux density from the NVSS
(column 3, Table 1) is a measure of the compactness of the FIRST
sources.  This compactness ratio is plotted in Figure~2 as a function
of the NVSS flux density, with plot symbols encoding the
identification status for the FIRST sources (column~4, Table~1).  The
FIRST and NVSS observations occured in 1994 July and 1995 April,
respectively.  In the absence of source variability, all sources are
expected to be located within the region defined by the dashed lines.
Figure~2 shows that the compactness ratios span the full range of
possibilites arising from source-resolution effects (from zero to
unity), while the few instances of ratios significantly exceeding
unity could plausibly be attributed to source variability.  Figure~2
also shows that the unidentified sources are spread throughout a wide
range of compactness ratios.

\section{VLBA Observations and Calibration}

The VLBA was used to observe the 76 FIRST sources and calibrators
during four observing segments described in Table~2.  For each
segment, antenna separations spanned 240 km to 8600 km.  Data were
acquired during about 6.5 hours in dual circular polarizations with
4-level sampling and at a center frequency 4.97949~GHz with a
bandwidth of 32~MHz per polarization.  This bandwidth was synthesized
from 4 contiguous baseband channels, each of width 8~MHz.  A strong
calibrator 3C\,279 was observed to align the phases of the independent
baseband channels.  Observation and correlation assumed a coordinate
equinox of 2000.  Correlation also assumed an integration time of 2
seconds, 32 spectral points per baseband channel, and a single pass at
the {\em a priori\/} position of each FIRST source.  While there were
a few instances where several FIRST sources fell within the primary
beam of a VLBA antenna, correlation was done only for the FIRST source
at the peak of the primary beam.

Phase-referenced observations were made in the nodding style at
elevations above 20\arcdeg.  Successive 80-second observations of
three FIRST sources were preceded and followed by a 60-second
observation of the phase, rate, and delay calibrator (referred to as a
phase calibrator hereafter) listed in Table~2, leading to a switching
time of 5~minutes.  {\em A priori\/} positions for the phase
calibrators were available in the International Celestial Reference
Frame (Extension 1) \citep[ICRF-Ext.1;][]{ier99,ma98}, so all FIRST
sources detected with the VLBA will have their positions measured in
that frame.  Each FIRST source was observed during six snapshots
spread over time to enhance coverage in the $(u,v)\/$ plane.  For each
segment, a maximum switching angle of 2.5\arcdeg\, was used to assign
FIRST sources to the segment's phase calibrator.  As is evident from
Figure~1, this procedure failed for one source (FIRST
J143152.544$+$340110.15), for which a maximum switching angle of
2.6\arcdeg\, was required.  Two phase calibrators were used during
segment 3 to reach different FIRST sources, so that segment is
conceptually split into segments 3a and 3b depending on which phase
calibrator was used (Table~2).  To help assess repeatability, some
FIRST sources were observed during two segments.  An observing
frequency of 5.0~GHz was used to optimize the chances that
phase-referencing would succeed for all switching angles;
phase-referencing at lower (higher) frequencies can be harmed by
ionospheric (tropospheric) effects \citep{wro00}.

Figure~1 and Table~2 also identify check calibrators that were
observed similarly to each FIRST source but with the switching angles
indicated.  Those calibrator observations were intended to serve two
purposes.  First, the tabulated positions are accurate to better than
1.3~mas per coordinate in the ICRF-Ext.1 and can be used to check the
differential astrometry.  Second, the calibrator data can be used to
check the coherence losses.  Phase calibrators for some segments
served as check calibrators for other segments.  Check calibrators
J1416$+$3444, F14543631, and F14653289 were also taken from the
table's references.  The latter two check calibrators were originally
selected from \citet{mye03} to improve upon the calibrator density
offered in \citet{bea02}.

For each segment, data editing and calibration were done with the NRAO
AIPS software and following the strategies outlined in Appendix C of
the NRAO AIPS
Cookbook\footnote{http://www.aoc.nrao.edu/aips/cook.html}.  Data
deletion was based on system flags recorded at observation and tape
weights recorded at correlation.  Corrections for the dispersive
delays caused by the Earth's ionosphere were made using
electron-content models based on Global Positioning System data and
derived at the Jet Propulsion Laboratory.  VLBA system temperatures
and gains were used to set the amplitude scale to an accuracy of about
5\%, after first correcting for sampler errors.  The visibility data
for the phase calibrator were used to generate phase-referenced
visibility data for the appropriate check calibrators (Table~2) and
for the subset of FIRST sources observed during that segment
(Table~1).  In contrast, the visibility data for the phase calibrator
and the strong calibrator 3C\,279 were self-calibrated.

Parallel and cross-hand correlations were obtained for each segment.
A polarization calibration was done for the first segment, but each
FIRST source was so faint in Stokes $I\/$ that Stokes $Q\/$ and $U\/$
were emission-free and did not usefully constrain the linear
polarization percentage $100 \sqrt{Q^2 + U^2} / I$.  Thus for later
segments a polarization calibration was not attempted.  For all
segments, however, the emission-free information in Stokes $Q\/$ for
each FIRST source proved to be a useful cross-check of the detection
threshold adopted in Stokes $I\/$.

\section{VLBA Imaging Strategies}

The AIPS task IMAGR was used to image the Stokes $I\/$ and $Q\/$
emission from each FIRST source and each calibrator.  To reduce
side-lobe levels for these VLBA multi-snapshot data, the visibility
data were weighted uniformly with robustness 0.5 and a sensitivity
loss of about 10\% relative to natural weighting was incurred.  A
two-stage approach to the imaging was taken.

The first image, not cleaned, spanned 4096 $\times$ 0.35~mas in each
coordinate and had a typical angular resolution characterized as an
elliptical Gaussian with FWHM dimensions of 3.0~mas by 1.5~mas aligned
nearly north-south (referred to as a resolution of 2~mas hereafter).
This image is within the field-of-view limits set by time and
bandwidth averaging \citep{wro95}; the most constraining limit follows
from accepting, at the field edge, a 10\% drop in the peak amplitude
due to averaging over each 8-MHz baseband channel, resulting in an
elliptical field of view aligned nearly north-south with major and
minor axes of 3000~mas and 1500~mas, respectively.  A search $\sigma$
was obtained by forming a histogram of brightness variations from
pixel to pixel within the search region and evaluating the
root-mean-square of that histogram.  Given the array, observation, and
imaging parameters, a detection threshold of 6 $\sigma$ $\sim$ 2~mJy
was adopted within the search region, a square of side 1400~mas.
Values for the search $\sigma$ vary from image to image and appear in
column 6 of Table~1.  For Stokes $I\/$, the right ascension and
declination of the peak of that detection was recorded.  For Stokes
$Q\/$, the emission-free search region was analysed to cross-check and
validate the 6 $\sigma$ detection threshold for Stokes $I\/$.  For a
strong Stokes $I\/$ detection, the quoted value for the search
$\sigma$ is typically higher than expected from thermal sensitivity,
because the histogram for the search region is corrupted by side-lobes
from the strong detection (the high-valued pixels of a strong Stokes
$I\/$ detection were excluded from the histogram).  Analysis of this
first set of images resulted in VLBA detections of 30 of the FIRST
sources in Table~1 and all calibrators.

A second image of Stokes $I\/$ was made for each VLBA detection,
including the 30 faint sources and all calibrators.  This second image
spanned 512 $\times$ 0.35~mas in each coordinate and was made by
shifting the tangent point, as derived from the right ascension and
declination of the peak in the first image, to the field center.
These shifted images were cleaned in regions centered on the
detections and spanning 2.1~mas east-west and 3.5~mas north-south for
the isolated detections but spanning customized regions for the
extended detections.  These shifts were finite for the
phase-referenced and cleaned images shown in Figures~3-5 but zero by
definition for the self-calibrated and cleaned images of the phase and
strong calibrators.

\subsection{Residual Phase Errors and VLBA Nondetections}

Residual errors during phase referencing will degrade the point-source
sensitivity of the VLBA.  To quantify the effect of this loss of
coherence, individual 80-second observations of the check calibrators
listed in Table~2 were phase self-calibrated, imaged, and cleaned.
The ratios of the peak intensities in the self-calibrated images were
1.0 to 1.4 times the peak intensities in the phase-referenced images,
as tabulated in column~10 of Table~2.  Observations of these
calibrators required switching angles of 0.6 to 2.7\arcdeg, but
switching angles of 2.6\arcdeg\, or less were needed to reach the
FIRST sources.  While there may be some segment-to-segment variations,
a conservative correction for the loss of point-source sensitivity is
about 1.2.  For the VLBA nondetections in Table~1, the quoted 6
$\sigma$ upper limits for a point source have been increased by this
factor.

\subsection{Modelfitting the VLBA Detections}

The bright components in the phase-referenced images (Figs.~3-5)
appear to be slightly resolved, but this apparent resolution is
probably mostly artificial and due to the residual phase errors
discussed in subsection 4.1.  In most cases, a single elliptical
Gaussian was fit to each of these VLBA images to yield the position
peak and integrated flux densities quoted in Table~1.  For a few
images, two elliptical Gaussians were fit and their integrated flux
densities were summed and tabulated along with the position of the
brighter Gaussian.  For the source showing very diffuse structure,
areal integration yielded the tabulated integrated flux density and a
parabolic fit of the brightest structure yielded the tabulated
position.  A few sources were too weak for reliable Gaussian fitting;
each of them is considered as a provisional VLBA detection, with a
parabolic fit yielding the tabulated peak flux density (corrected for
coherence losses) and position.

For the VLBA integrated flux density, no corrections were made for
coherence losses and the tabulated error is the quadratic sum of the
5\% scale error and the error in the Gaussian model.  Although the
phase errors leading to coherence loss do broaden the VLBA detections,
coherence corrections were not made because the flux densities can be
largely recovered with Gaussian fitting.  For the VLBA positions, the
calculated error per coordinate is the quadratic sum of three terms:
(a) the 1 $\sigma$ error in the phase calibrator position; (b) the 1
$\sigma$ error in the differential astrometry; and (c) the 1 $\sigma$
error in the Gaussian fit.  As indicated in the note to Table~2, term
(a) is always less than 1~mas in one dimension; to be conservative,
1~mas is adopted for term (a).  To quantify term (b), the positions of
the check calibrators in Table~2 were measured in the phase-referenced
images and subtracted from their {\em a priori\/} positions, leading
to the position corrections, in right ascension and declination,
tabulated in column~11 of Table~2.  Observations of these check
calibrators required switching angles of 0.6 to 2.7\arcdeg, but
switching angles of 2.6\arcdeg\, or less were needed to reach the
FIRST sources.  While there may be some segment-to-segment variations,
a conservative estimate for term (b) is 2~mas in one dimension.  For
the weakest VLBA detections of FIRST sources, term (c) has a
worst-case value of 0.3~mas in the north-south direction and half that
in the east-west direction; to be conservative, 1~mas is adopted for
term (c).  Thus for all the VLBA detections reported in Table~2, a
conservative estimate of the position error per coordinate is the
quadratic sum of 1~mas, 2~mas, and 1~mas, or 2.5~mas.

\section{Discussion}

\subsection{VLBA Overview}

Seventy-six faint radio sources were selected from the FIRST catalog
to have peak flux densities above 10~mJy at 5\arcsec\, resolution and
deconvolved major diameters of less than 3\arcsec\, at 1.4~GHz.  The
compactness ratio on VLA scales at 1.4~GHz, as defined in Section~3,
is plotted in Figure~6 as a function of the VLBA photometry at 5.0~GHz
(column~9, Table~1).  As for Figure~2, the plot symbols encode the
identification status for the FIRST sources (column~5, Table~1).
Figure~6 shows that VLBA detections occur over the full range of
compactness ratios.

\subsection{VLBA Detection Rate}

Thirty of the 76 FIRST sources, or 39$^{+9}_{-7}$\%, were detected
with the VLBA at 5.0~GHz with peak flux densities above 6 $\sigma
\sim$ 2~mJy at 2~mas resolution.  For 20 similar but somewhat fainter
sources in the {\em Spitzer\/} First-Look Survey \citep{con03}, twelve
sources, or 60$^{+23}_{-17}$\%, were detected with the VLBA at
1.4~GHz, with peak flux densities above 2~mJy at 9 mas resolution
\citep{wro04}.  Equally good success rates can be expected when VLBA
phase-referencing in the nodding style is used to target similar FIRST
sources anywhere in that survey's 10,000~deg$^2$.  Moreover, among the
VLBA detections reported in Table~1, FIRST J142910.223$+$352946.86 was
used successfully as an in-beam phase calibrator \citep{wro00} at
1.4~GHz for a VLBI survey of tens of sources covering 0.28~deg$^2$ in
the NOAO Bo\"otes Field \citep{gar05}.  This illustrates how the
stronger VLBA detections can, themselves, serve as VLBI phase
calibrators, potentially opening up much of the radio sky to
phase-referencing in either the nodding or the in-beam styles.

\subsection{VLBA Parsec-Scale Structures}

For the adopted cosmology, the achieved VLBA resolution of 2~mas
correponds to 17~pc or finer, while each of the panels in Figures~3-5
spans 170~pc or less.  Most VLBA detections at 5.0~GHz are unresolved
or appear to be slightly resolved, with rest-frame brightness
temperatures in the range $(2-50) (1+z) \times 10^7$~K, or more if
truly unresolved.  Such high brightness temperatures, in combination
with the optical identifications discussed below, imply that these
VLBA detections mark accretion-driven outflows on parsec scales.
Also, one VLBA detection (FIRST J142905.105$+$342641.06) appears quite
diffuse and elongated, as discussed in the appendix.  Five others
(FIRST J143121.320$+$332808.95, FIRST J143152.544$+$340110.15, FIRST
J143449.111$+$354246.98, FIRST J143752.050$+$351940.08, and FIRST
J143841.949$+$335809.48) show either double or core-jet structures, as
mentioned in the appendix.  The structures of these six VLBA
detections further enforce an outflow scenario, and we now consider
some issues that can be addressed based on these structures.

Could some of the five VLBA detections showing double or core-jet
structures eventually yield evidence for counterjets, important for
strong tests of unified schemes \citep{xu00}?  For sources selected
independent of their radio spectra, more counterjets can be expected
adjacent to core-jet structures, but may elude discovery until
follow-up VLBA observations with higher sensitivity.

Might some of the five VLBA detections showing double or core-jet
structures be compact symmetric objects (CSOs), young systems offering
insights into evolutionary models for radio galaxies and strong tests
of unified schemes \citep{tay00}?  From VLBI imaging of a bright and
flat-spectrum sample \citep{tay96}, \citet{tay00} note that about 7\%
show CSO structures.  Using this as a rough guide for faint sources
selected independent of their radio spectra, two CSO candidates can be
expected among the 30 VLBA detections.  But for sources selected
independent of their radio spectra, more symmetric core-jet structures
are to be expected, giving them the appearance of CSOs in the
discovery images.  Thus follow-up VLBA observations at other
frequencies will be required to distinquish between candidate
core-jets and candidate CSOs.

Finally, is it likely that some of the five VLBA detections showing
double or core-jet structures could be gravitational millilenses,
caused by a cosmic population of supermassive black holes
\citep{wil01}?  From VLBI imaging of a bright and flat-spectrum sample
\citep{tay96}, \citet{wil01} find that for image separations
1.5-50~mas, the upper limit at 95\%-confidence to the expected
millilensing rate is about one lens per 430 sources.  Using this as a
rough guide for faint sources selected independent of their radio
spectra, then two implications follow.  First, the prospect of
discovering one millilens candidate among 30 VLBA detections is slim.
Second, expanding the present VLBA survey by a factor of 14 would lead
to equally good constraints on the expected millilensing rate.  More
than a thousand sources would have been examined in the process, with
about four in ten being VLBA detections searched for lensing
signatures.  This VLBA survey method uses 0.5 hours to observe three
sources, so repeating the survey in 13 other regions would require
additional observations totaling 164 hours.

\subsection{VLBA Detections Unidentifed in the NOAO Bo\"otes Survey}

Fifty-seven of the 70 FIRST sources, or 81\%, were identified with
accretion-powered radio galaxies and quasars using the NOAO data
\citep{jan05}.  A similar identification percentage applies to the 27
VLBA detections, as 24, or 89\%, were identified.

The three VLBA detections with fainter optical hosts must still mark
active nuclei.  One of these three is an apparent double with an 8-mas
separation (FIRST J143121.320$+$332808.95), while the other two (FIRST
J142738.625$+$330756.96 and FIRST J143643.209$+$352222.98) are
unresolved or appear to be slightly resolved.  These three
unidentified VLBA detections at 5.0~GHz have rest-frame brightness
temperatures in the range $(2-6) (1+z) \times 10^7$~K, or more if
truly unresolved.  This is similar to the range of $(2-50) (1+z)
\times 10^7$~K for the identified VLBA detections that pin-point the
active nuclei of the accretion-powered radio galaxies and quasars.

The three unidentified VLBA detections fainter than 25.5~mag in the
$I\/$ band are likely candidates for being optically-obscured active
nuclei at high redshift.  Also in the NOAO Bo\"otes field,
\citet{gar05} recognized a similar case involving a VLBI detection at
1.4~GHz of an unidentified 20-mJy Westerbork source, while
\citet{hig05} report numerous cases of unidentified VLA sources in the
mJy regime at 1.4~GHz, for which analysis of their spectral energy
distributions (SEDs) supports an accretion-power origin.  Rector et
al.\ (2005, in preparation) will investigate the SEDs of the three
unidentified VLBA detections mentioned above, paying particular
attention to a comparison of their SEDs to those of the unidentified
VLA sources analyzed by \citet{hig05}.

\section{Summary and Future Steps}

The VLBA was used at 5.0~GHz to obtain phase-referenced images of 76
sources in the NOAO Bo\"otes field, selected from the FIRST catalog to
have peak flux densities above 10~mJy at 5\arcsec\, resolution and
deconvolved major diameters of less than 3\arcsec\, at 1.4~GHz.
Fifty-seven of these faint radio sources were identified in the
optical with accretion-powered radio galaxies and quasars with $I\/ <
25.5$~mag.  On VLA scales at 1.4~GHz, a measure of the compactness of
the faint sources is the ratio of the peak flux density from FIRST to
the integrated flux density from the NVSS catalog.  These compactness
ratios span the full range of possibilites arising from
source-resolution effects.  Also, the unidentified sources are spread
throughout a wide range of compactness ratios.

Thirty of the faint radio sources (39$^{+9}_{-7}$\%) were detected
with the VLBA at 5.0~GHz, with peak flux densities above 6 $\sigma
\sim$ 2~mJy at 2~mas resolution.  The VLBA detections occur through
the full range of compactness ratios.  For the adopted cosmology, the
VLBA resolution correponds to 17~pc or finer.  Most VLBA detections
are unresolved or slightly resolved.  The resulting high brightness
temperatures, in combination with the optical identifications, imply
that these VLBA detections mark accretion-driven outflows on parsec
scales.  Moreover, the elongated structures of six VLBA detections
further enforce an outflow scenario.  Three VLBA detections are
unidentified and fainter than 25.5~mag in the $I\/$ band, similar to
another VLBI detection reported by \citet{gar05}.  These four VLBI
detections are likely candidates for being optically-obscured active
nuclei at high redshift, so understanding them is essential for
completing the cosmic census of supermassive black holes.

For these faint radio sources in the NOAO Bo\"otes field, the next
steps involve obtaining redshifts, assessing intrinsic properties,
examining the properties of the optical counterparts and their cluster
environments, assembling the SEDs of the active nuclei, and tying data
for the Bo\"otes field to the ICRF.  These topics will be covered by
Rector et al.\ (2005, in preparation).  Here we simply note that
concerning the SEDs, for the 30 VLBA detections in Table~1, the
photometry quantifies the emission from the active nuclei at 5.0~GHz;
while for the 46 VLBA nondetections in Table~1, the photometry imposes
upper limits on the emission from the active nuclei at 5.0~GHz.
Rector et al.\ (2005, in preparation) will present further constraints
on the SEDs of the active nuclei from the NOAO \citep{jan05, dey05}
and {\em Spitzer\/} photometry \citep{eis04, hou05}.  In Figure~1, the
crosses in the north-east corner of the Bo\"otes field show the four
X-ray sources from \citet{wan04} with FIRST counterparts;
unfortunately, all are weaker than the 10~mJy threshold adopted for
this study.  An X-ray survey covering more of the Bo\"otes field is
clearly needed.  Finally, we note that concerning the ICRF,
twenty-four of the VLBA detections are identified (column~5, Table~1,
Fig.~6), so the VLBA astrometry will help anchor the multiwavelength
data for the NOAO Bo\"otes field to the ICRF.

\acknowledgments 

We acknowledge the anonymous referee for suggestions that helped
clarify the manuscript and Ian McGreer for identification feedback.
The calibrator source positions listed in Table~2 in the frame of the
ICRF-Ext.1 catalog were provided by observations from the joint
NASA/USNO/NRAO geodetic/astrometric program.  This work made use of
images and/or data products provided by the NOAO Deep Wide-Field
Survey \citep{jan99, jan05, dey05}, which is supported by the National
Optical Astronomy Observatory (NOAO).  NOAO is operated by AURA, Inc.,
under a cooperative agreement with the National Science Foundation.

\appendix

\section{Notes on Individual Sources}

This appendix contains notes on the sources flagged in Table~1.  An
NVSS secondary is noted if present with an offset of less than
90\arcsec.  A FIRST secondary is noted if present within a square of
side 90\arcsec\, centered on the FIRST position.  A note appears for
all VLBA detections.

{\bf FIRST J142456.287$+$352841.80:} An NVSS secondary is also
present, with $S_{\rm I}=16.7\pm0.6$~mJy offset by 68\arcsec at PA
83\arcdeg.  The FIRST image shows secondaries to the east and west of
the FIRST position.  The VLBA images in Fig.~3 from segments 1 and 2
have 1 $\sigma$ noise levels of 0.23 and 0.30 mJy~beam$^{-1}$,
respectively.  Possibly a photometric variable at 5.0~GHz at 2~mas
resolution.  There is a counterpart in the optical $I\/$ band so this
VLBA detection will help link data for the Bo\"otes field to the ICRF.

{\bf FIRST J142524.214$+$340935.68:} The VLBA image in Fig.~3 has a 1
$\sigma$ noise level of 0.32 mJy~beam$^{-1}$.  There is a counterpart
in the optical $I\/$ band so this VLBA detection will help link data
for the Bo\"otes field to the ICRF.

{\bf FIRST J142607.716$+$340426.29:} The VLBA image in Fig.~3 has a 1
$\sigma$ noise level of 0.34 mJy~beam$^{-1}$.  There is a counterpart
in the optical $I\/$ band so this VLBA detection will help link data
for the Bo\"otes field to the ICRF.

{\bf FIRST J142617.948$+$344039.63, FIRST J142623.374$+$343950.45:}
These FIRST sources are part of the same NVSS double.  FIRST
J142623.374$+$343950.45 is identifed in the optical $I\/$ band and
appears isolated in the FIRST image.  FIRST J142623.374$+$343950.45 is
unidentifed in the optical $I\/$ band and shows a secondary in the
north-west quadrant of the FIRST image; perhaps it is a lobe hotspot
energized by the active nucleus in the optical host for FIRST
J142623.374$+$343950.45.

{\bf FIRST J142659.719$+$341200.21:} The FIRST source is the central
component of an NVSS double with the tabulated $S_{\rm I}$.  The NVSS
double shows Fanaroff-Riley I structure in the north-east and
south-west quadrants of the FIRST image.

{\bf FIRST J142738.625$+$330756.96, FIRST J142739.694$+$330744.45:}
These FIRST sources are part of the same NVSS source, so they share
the same tabulated $S_{\rm I}$ from NVSS.  FIRST
J142738.625$+$330756.96 is provisionally detected with the VLBA and is
unidentifed in the optical $I\/$ band; its VLBA image shown in Fig.~3
has a 1 $\sigma$ noise level of 0.27 mJy~beam$^{-1}$.  FIRST
J142739.694$+$330744.45 is undetected with the VLBA and is identified
in the optical $I\/$ band.

{\bf FIRST J142744.441$+$333828.62:} The VLBA image in Fig.~3 has a 1
$\sigma$ noise level of 0.37 mJy~beam$^{-1}$.  There is a counterpart
in the optical $I\/$ band so this VLBA detection will help link data
for the Bo\"otes field to the ICRF.

{\bf FIRST J142758.722$+$324741.56:} The FIRST source is the central
component of an NVSS double with the tabulated $S_{\rm I}$.  This NVSS
double shows an east-west, Fanaroff-Riley II structure in the FIRST
image.  The VLBA image in Fig.~3 has a 1 $\sigma$ noise level of 0.43
mJy~beam$^{-1}$.  There is a counterpart in the optical $I\/$ band so
this VLBA detection will help link data for the Bo\"otes field to the
ICRF.

{\bf FIRST J142806.696$+$325935.83:} The VLBA image in Fig.~3 has a 1
$\sigma$ noise level of 0.53 mJy~beam$^{-1}$.  There is a counterpart
in the optical $I\/$ band so this VLBA detection will help link data
for the Bo\"otes field to the ICRF.

{\bf FIRST J142842.556$+$354326.60:} The VLBA images in Fig.~3 from
segments 1 and 2 have 1 $\sigma$ noise levels of 0.24 and 0.30
mJy~beam$^{-1}$, respectively.  There is a counterpart in the optical
$I\/$ band so this VLBA detection will help link data for the Bo\"otes
field to the ICRF.

{\bf FIRST J142904.545$+$343243.54:} The NVSS catalog notes a complex
source structure.  The NVSS image shows a weak extension to the
north-east, likely related to a weak secondary in the FIRST image in
the same quadrant.

{\bf FIRST J142905.105$+$342641.06:} The VLBA images in Fig.~3 from
segments 2 and 4 have 1 $\sigma$ noise levels of 0.31 and 0.29
mJy~beam$^{-1}$, respectively.  The integrated flux density and the
diffuse structure are repeatable between the two segments, but the
VLBA image fidelity is likely to be quite poor.  On VLA scales at
1.4~GHz, this is among the strongest FIRST sources and its compactness
ratio is unity.  There is a counterpart in the optical $I\/$ band so
this VLBA detection will help link data for the Bo\"otes field to the
ICRF.

{\bf FIRST J142910.223$+$352946.86:} The VLBA images in Fig.~4 from
segments 1 and 2 have 1 $\sigma$ noise levels of 0.44 and 0.35
mJy~beam$^{-1}$, respectively.  \citet{gar05} present a VLBI image at
1.4~GHz.  There is a counterpart in the optical $I\/$ band so this
VLBA/VLBI detection will help link data for the Bo\"otes field to the
ICRF.

{\bf FIRST J142911.747$+$333144.24:} The FIRST image shows a weak
secondary to the south of the FIRST position.

{\bf FIRST J142937.566$+$344115.69:} An NVSS secondary is also
present, with $S_{\rm I}=3.4\pm0.5$~mJy offset by 86\arcsec at PA
$-$90\arcdeg, perhaps related to a weak secondary in the FIRST image
in a similar PA.  The VLBA image in Fig.~4 has a 1 $\sigma$ noise
level of 0.26 mJy~beam$^{-1}$ and is the basis for the provisional
VLBA detection from segment 1.  The less sensitive search from segment
2 resulted in a nondetection.  There is a counterpart in the optical
$I\/$ band so this provisional VLBA detection might help link data for
the Bo\"otes field to the ICRF.

{\bf FIRST J143022.337$+$343727.14:} The FIRST image shows a secondary
to the east of the FIRST position.

{\bf FIRST J143056.092$+$341930.18:} The FIRST image shows a secondary
in the north-west quadrant.

{\bf FIRST J143109.858$+$335301.67:} An NVSS secondary is also
present, with $S_{\rm I}=2.3\pm0.4$~mJy offset by 82\arcsec at PA
103\arcdeg.

{\bf FIRST J143121.320$+$332808.95:} The VLBA image in Fig.~4 has a 1
$\sigma$ noise level of 0.30 mJy~beam$^{-1}$.  This image shows either
a double or a core-jet structure.  There is no counterpart in the
optical $I\/$ band.

{\bf FIRST J143123.297$+$331625.82:} The VLBA image in Fig.~4 has a 1
$\sigma$ noise level of 0.30 mJy~beam$^{-1}$ and is the basis for the
provisional VLBA detection.  There is no counterpart in the optical
$I\/$ band catalog but the $I\/$ band image shows an object 1\arcsec\,
to the east.

{\bf FIRST J143134.549$+$351511.19:} The NVSS catalog notes a complex
source structure.  The NVSS image shows a weak extension to the
south-east, likely related to a weak secondary in the FIRST image in
the same quadrant.

{\bf FIRST J143152.544$+$340110.15:} The VLBA images in Fig.~4 from
segments 2 and 4 have 1 $\sigma$ noise levels of 0.30 and 0.27
mJy~beam$^{-1}$, respectively.  The integrated flux density and the
double or core-jet structure are repeatable between the two segments.
There is a counterpart in the optical $I\/$ band so this VLBA
detection will help link data for the Bo\"otes field to the ICRF.

{\bf FIRST J143309.671$+$351520.14:} The FIRST image shows a secondary
in the south-east quadrant.

{\bf FIRST J143311.054$+$335828.81:} The VLBA image in Fig.~4 has a 1
$\sigma$ noise level of 0.23 mJy~beam$^{-1}$.  There is a counterpart
in the optical $I\/$ band so this VLBA detection will help link data
for the Bo\"otes field to the ICRF.

{\bf FIRST J143345.947$+$341149.03:} An NVSS secondary is also
present, with $S_{\rm I}=100.2\pm3.6$~mJy offset by 77\arcsec at PA
$-$102\arcdeg, perhaps related to a secondary in the south-west
quadrant of the FIRST image.  There is no counterpart in the optical
$I\/$ band so FIRST J143345.947$+$341149.03 might be a lobe hotspot.

{\bf FIRST J143432.836$+$352141.47:} The FIRST image shows a secondary
in the south-east quadrant.

{\bf FIRST J143434.217$+$351009.53:} The VLBA image in Fig.~4 has a 1
$\sigma$ noise level of 0.69 mJy~beam$^{-1}$.  There is a counterpart
in the optical $I\/$ band so this VLBA detection will help link data
for the Bo\"otes field to the ICRF.

{\bf FIRST J143445.321$+$332820.58:} The VLBA image in Fig.~4 has a 1
$\sigma$ noise level of 0.27 mJy~beam$^{-1}$.  It was formed from
phase self-calibrated data, yielded the tabulated VLBA flux density,
and was used to estimate a coherence correction for segment 4.  The
phase-referenced image yielded the tabulated VLBA position and was
used to estimate a coherence correction for segment 4.  There is a
counterpart in the optical $I\/$ band so this VLBA detection will help
link data for the Bo\"otes field to the ICRF.

{\bf FIRST J143446.491$+$332452.27:} The VLBA image in Fig.~4 has a 1
$\sigma$ noise level of 0.25 mJy~beam$^{-1}$ and is the basis for the
provisional VLBA detection.  There is a counterpart in the optical
$I\/$ band so this provisional VLBA detection might help link data for
the Bo\"otes field to the ICRF.

{\bf FIRST J143449.111$+$354246.98:} The NVSS image shows an asymmetry
to the east, seemingly related to the eastern secondary in the FIRST
image.  The VLBA image in Fig.~4 has a 1 $\sigma$ noise level of 0.38
mJy~beam$^{-1}$ and in Fig.~5 has a 1 $\sigma$ noise level of 0.30
mJy~beam$^{-1}$.  The integrated flux density and the double or
core-jet structure are repeatable between the two segments.  There is
a counterpart in the optical $I\/$ band so this VLBA detection will
help link data for the Bo\"otes field to the ICRF.

{\bf FIRST J143527.951$+$331145.46:} The NVSS image shows an asymmetry
to the south-east, seemingly related to the weak secondary in the same
quadrant of the FIRST image.  The VLBA image in Fig.~5 has a 1
$\sigma$ noise level of 0.23 mJy~beam$^{-1}$ and is the basis for the
provisional VLBA detection.  There is a counterpart in the optical
$I\/$ band so this provisional VLBA detection might help link data for
the Bo\"otes field to the ICRF.

{\bf FIRST J143528.374$+$331931.53:} The VLBA image in Fig.~5 has a 1
$\sigma$ noise level of 0.36 mJy~beam$^{-1}$.  There is a counterpart
in the optical $I\/$ band so this VLBA detection will help link data
for the Bo\"otes field to the ICRF.

{\bf FIRST J143529.161$+$343422.94:} The VLBA image in Fig.~5 has a 1
$\sigma$ noise level of 0.24 mJy~beam$^{-1}$.  There is a counterpart
in the optical $I\/$ band so this VLBA detection will help link data
for the Bo\"otes field to the ICRF.

{\bf FIRST J143543.629$+$353305.02:} The FIRST image shows a secondary
in the north-west quadrant.

{\bf FIRST J143643.209$+$352222.98:} The VLBA image in Fig.~5 has a 1
$\sigma$ noise level of 0.32 mJy~beam$^{-1}$.  There is no counterpart
in the optical $I\/$ band.

{\bf FIRST J143647.710$+$331037.66:} The FIRST image shows a secondary
to the south.

{\bf FIRST J143713.571$+$350554.84:} The VLBA image in Fig.~5 has a 1
$\sigma$ noise level of 0.54 mJy~beam$^{-1}$.  There is a counterpart
in the optical $I\/$ band so this VLBA detection will help link data
for the Bo\"otes field to the ICRF.

{\bf FIRST J143723.715$+$350734.69:} The NVSS catalog notes a complex
source structure.  The NVSS image shows this is likely due to the
proximity of an NVSS secondary, cataloged as having $S_{\rm
I}=14.6\pm0.6$~mJy offset by 70\arcsec at PA 78\arcdeg.

{\bf FIRST J143728.413$+$331110.20:} An NVSS secondary is also
present, with $S_{\rm I}=8.2\pm0.5$~mJy offset by 77\arcsec at PA
23\arcdeg.  The VLBA image in Fig.~5 has a 1 $\sigma$ noise level of
0.31 mJy~beam$^{-1}$.

{\bf FIRST J143752.050$+$351940.08:} The NVSS catalog notes a complex
source structure.  The NVSS image shows this is likely due to the
proximity of an NVSS secondary, cataloged as having $S_{\rm
I}=96.8\pm3.8$~mJy offset by 53\arcsec at PA 90\arcdeg.  The VLBA
image in Fig.~5 has a 1 $\sigma$ noise level of 0.28 mJy~beam$^{-1}$
and shows either a double or a core-jet structure.  There is a
counterpart in the optical $I\/$ band so this VLBA detection will help
link data for the Bo\"otes field to the ICRF.

{\bf FIRST J143841.949$+$335809.48:} The VLBA image in Fig.~5 has a 1
$\sigma$ noise level of 0.30 mJy~beam$^{-1}$ and shows either a double
or a core-jet structure.  There is a counterpart in the optical $I\/$
band so this VLBA detection will help link data for the Bo\"otes field
to the ICRF.

{\bf FIRST J143850.267$+$340419.84:} The VLBA image in Fig.~5 has a 1
$\sigma$ noise level of 0.28 mJy~beam$^{-1}$.  The identification
status is based on the $R\/$ band, as the $I\/$ band data are
corrupted or missing.

{\bf FIRST J143859.469$+$345309.33:} The FIRST image shows a secondary
in the north-east quadrant.  The identification status is based on the
$R\/$ band, as the $I\/$ band data are corrupted or missing.

{\bf FIRST J143909.406$+$332101.73:} The FIRST source appears to be
associated with the stronger component of the compact NVSS double with
the tabulated $S_{\rm I}$.  The NVSS catalog notes a complex source
structure for both components of the double, but areal integration
over the NVSS image confirms the tabulated $S_{\rm I}$.  The VLBA
image in Fig.~5 has a 1 $\sigma$ noise level of 0.34 mJy~beam$^{-1}$.

{\bf FIRST J143916.249$+$344912.09:} The VLBA image in Fig.~5 has a 1
$\sigma$ noise level of 0.34 mJy~beam$^{-1}$.

\clearpage

\begin{figure}
\epsscale{0.85}
\plotone{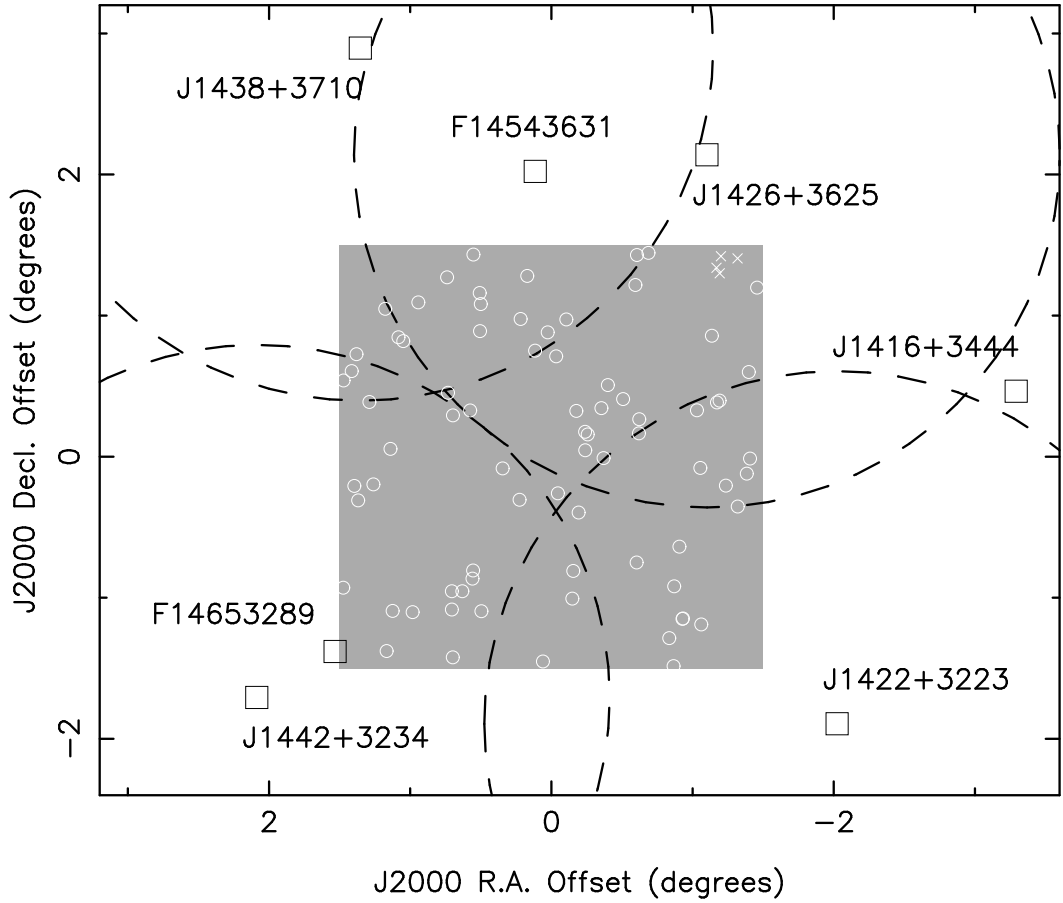}
\caption{Geometry for the VLBA observations at 5.0~GHz within the NOAO
Bo\"otes field, shown as a grey square covering 9.0~deg$^2$.  Light
circles show the locations of the FIRST sources listed in Table~1.
Dark squares show the locations of calibrator sources, four of which
served as phase calibrators for the VLBA observations of FIRST sources
within the large dashed circles, each of radius 2.5\arcdeg .  Crosses
in north-east corner of the Bo\"otes field show the four X-ray sources
from Wang et al.\ with FIRST counterparts, all weaker than 10~mJy.}
\end{figure}

\clearpage

\begin{figure}
\includegraphics[scale=.65,angle=-90]{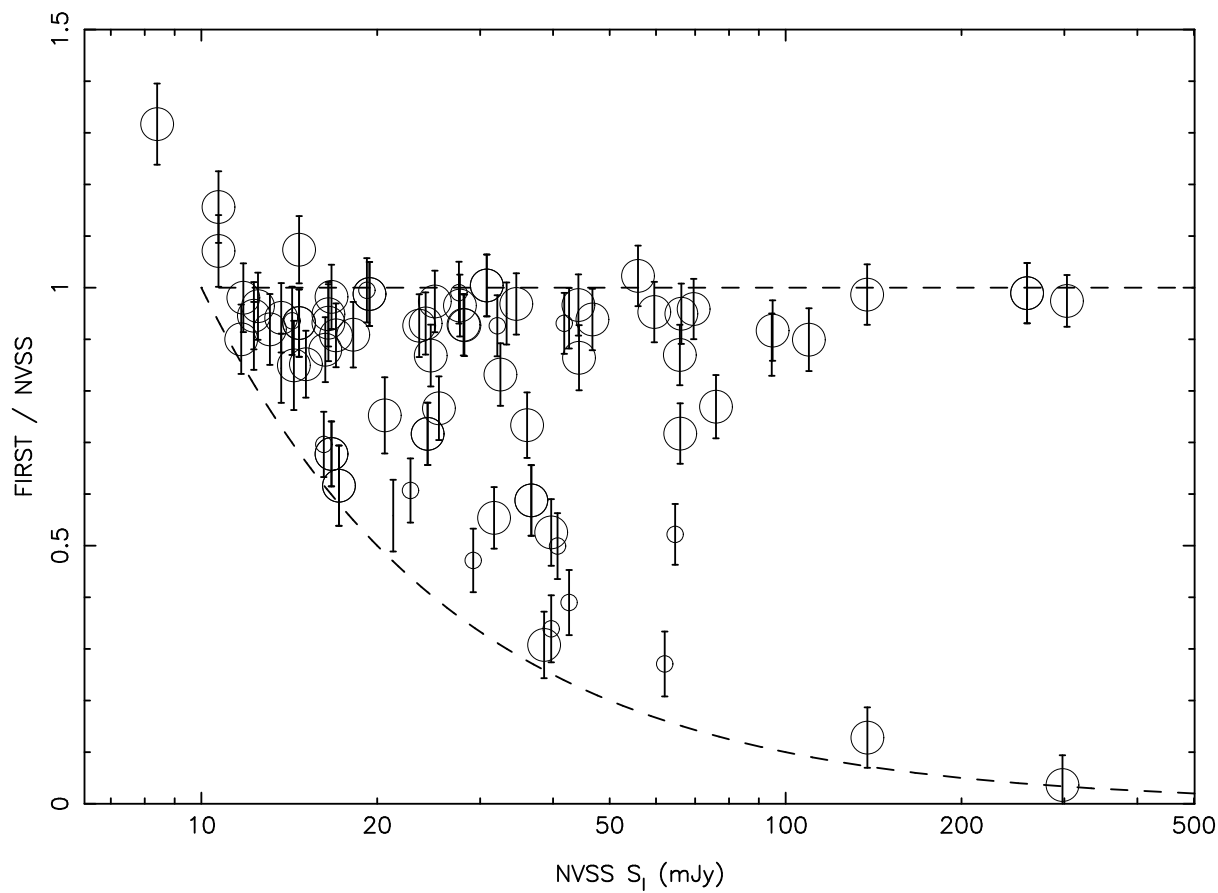}
\caption{Ordinate is the ratio at 1.4~GHz of the peak flux density at
5\arcsec\, resolution from FIRST to the integrated flux density at
45\arcsec\, resolution from the NVSS.  This ratio is a measure of the
compactness of the FIRST source.  Abscissa is the integrated NVSS flux
density.  The dashed horizontal line corresponds to a source whose
peak FIRST flux density and integrated NVSS flux density are equal,
and have a value of 10 mJy or more.  The dashed curved line
corresponds to a source whose peak FIRST flux density is 10 mJy, while
its integrated NVSS flux density is 10 mJy or more.  A large circle
means the FIRST source is identified in the optical $I\/$ band, while
a small circle means it is unidentified.}
\end{figure}

\clearpage

\begin{figure}
\epsscale{0.7}
\plotone{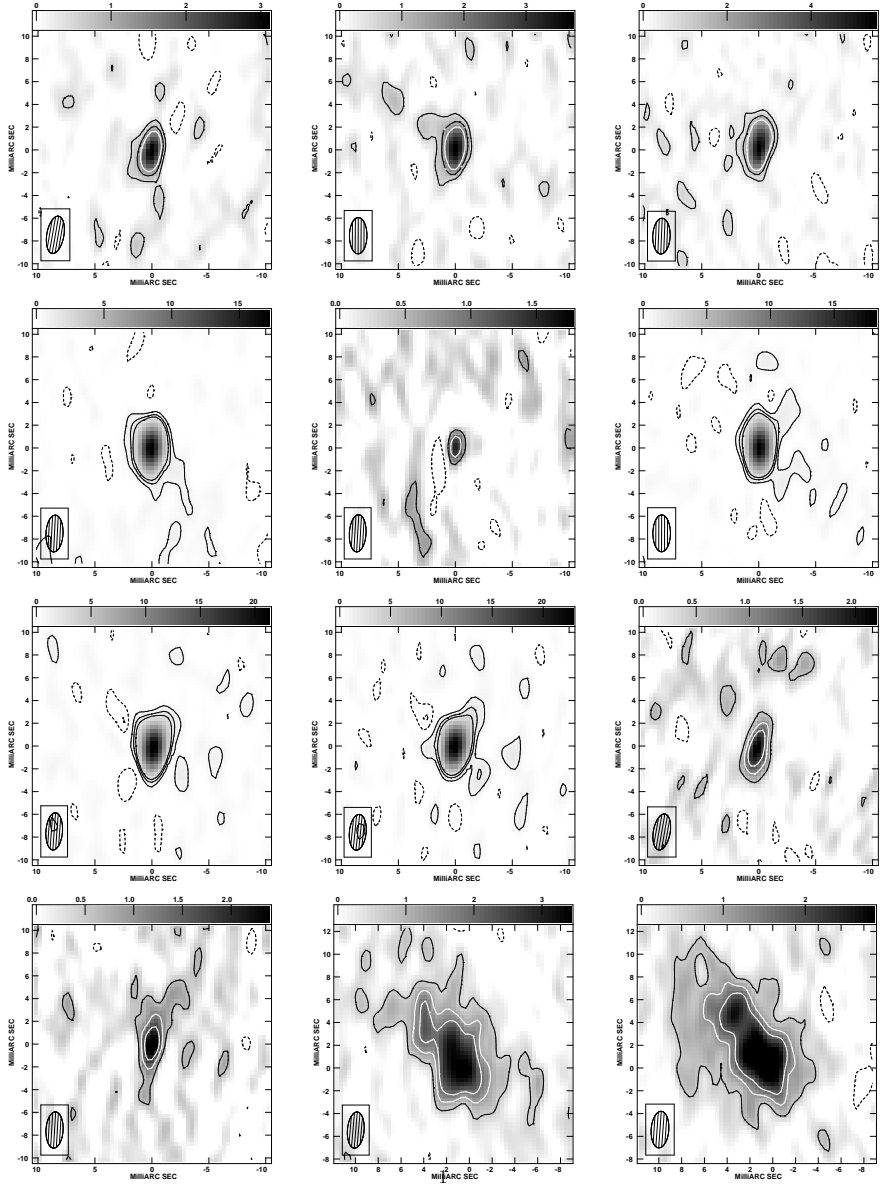}
\caption{Phase-referenced and cleaned images at 5.0~GHz of Stokes
$I\/$ emission for FIRST sources detected with the NRAO VLBA within a
square search region of side 1400~mas.  The boxed ellipse shows the
Gaussian restoring beam at FWHM.  Contours are at $\pm$2, $\pm$4, and
$\pm$6 times the 1 $\sigma$ noise level quoted in the appendix.
{\em Left to right, from the top:\/} 
FIRST J142456.287$+$352841.80 (segment 1),
FIRST J142456.287$+$352841.80 (segment 2),
FIRST J142524.214$+$340935.68, 
FIRST J142607.716$+$340426.29,
FIRST J142738.625$+$330756.96, 
FIRST J142744.441$+$333828.62,
FIRST J142758.722$+$324741.56, 
FIRST J142806.696$+$325935.83,
FIRST J142842.556$+$354326.60 (segment 1),
FIRST J142842.556$+$354326.60 (segment 2),
FIRST J142905.105$+$342641.06 (segment 2),
FIRST J142905.105$+$342641.06 (segment 4)
.}
\end{figure}

\clearpage

\begin{figure}
\epsscale{0.8}
\plotone{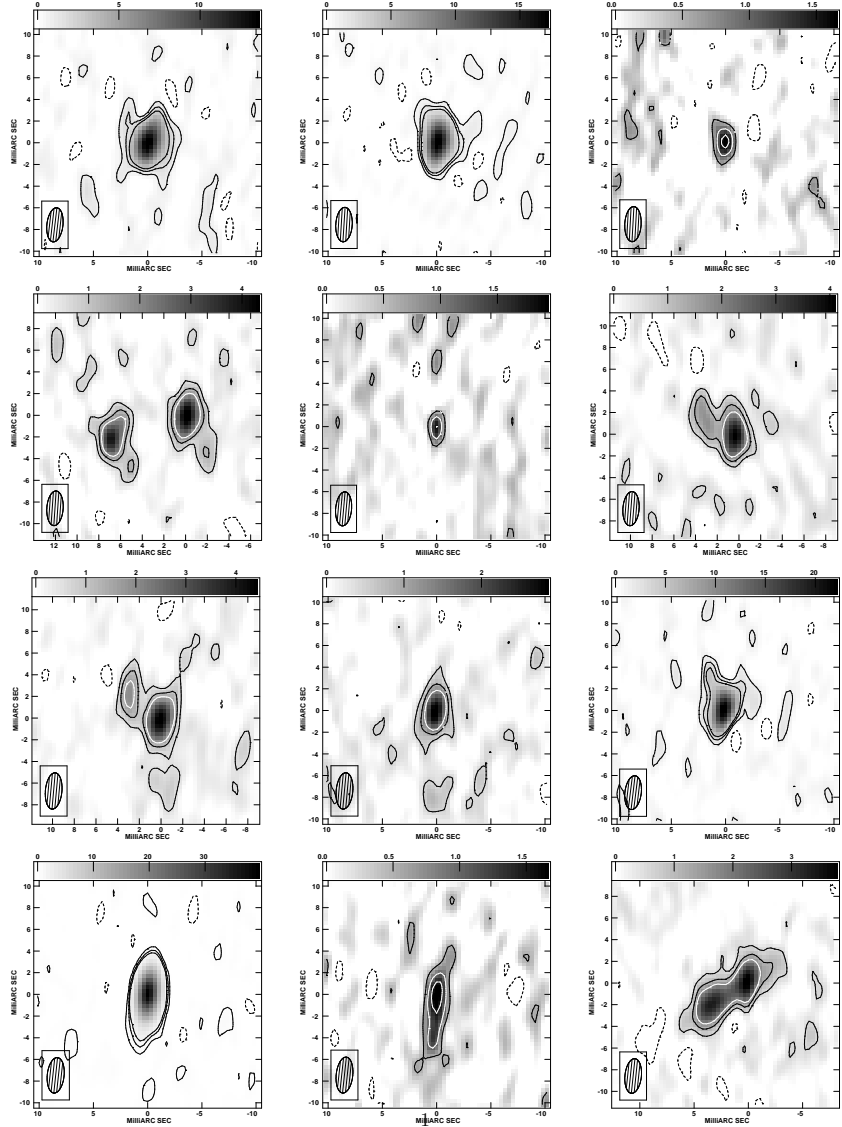}
\caption{Same as Fig.~3, but for
FIRST J142910.223$+$352946.86 (segment 1),
FIRST J142910.223$+$352946.86 (segment 2),
FIRST J142937.566$+$344115.69, 
FIRST J143121.320$+$332808.95,
FIRST J143123.297$+$331625.82,
FIRST J143152.544$+$340110.15 (segment 2),
FIRST J143152.544$+$340110.15 (segment 4),
FIRST J143311.054$+$335828.81,
FIRST J143434.217$+$351009.53,
FIRST J143445.321$+$332820.58,
FIRST J143446.491$+$332452.27,
FIRST J143449.111$+$354246.98 (segment 1)
.}
\end{figure}

\clearpage

\begin{figure}
\epsscale{0.7}
\plotone{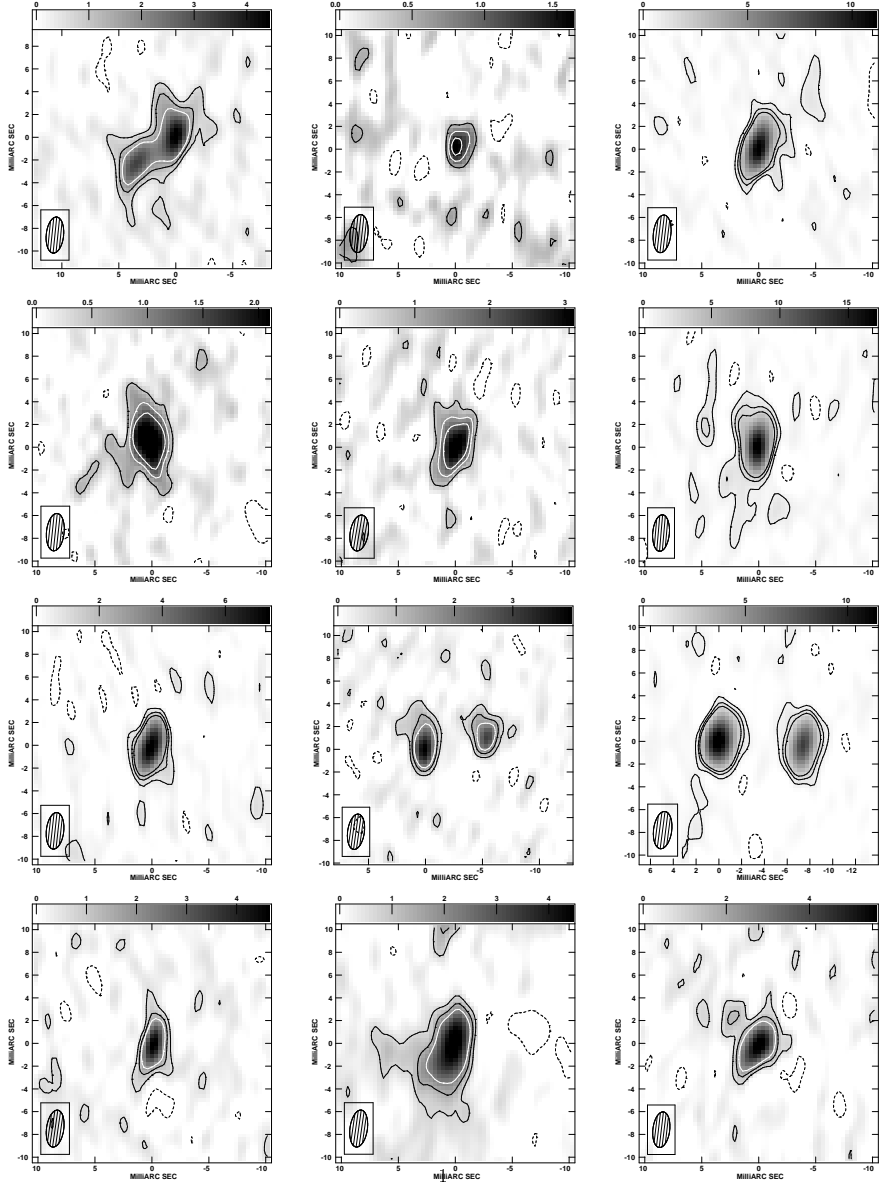}
\caption{Same as Fig.~3, but for
FIRST J143449.111$+$354246.98 (segment 2),
FIRST J143527.951$+$331145.46,
FIRST J143528.374$+$331931.53,
FIRST J143529.161$+$343422.94,
FIRST J143643.209$+$352222.98,
FIRST J143713.571$+$350554.84,
FIRST J143728.413$+$331110.20,
FIRST J143752.050$+$351940.08,
FIRST J143841.949$+$335809.48,
FIRST J143850.267$+$340419.84,
FIRST J143909.406$+$332101.73,
FIRST J143916.249$+$344912.09
 .}
\end{figure}

\clearpage

\begin{figure}
\includegraphics[scale=.65,angle=-90]{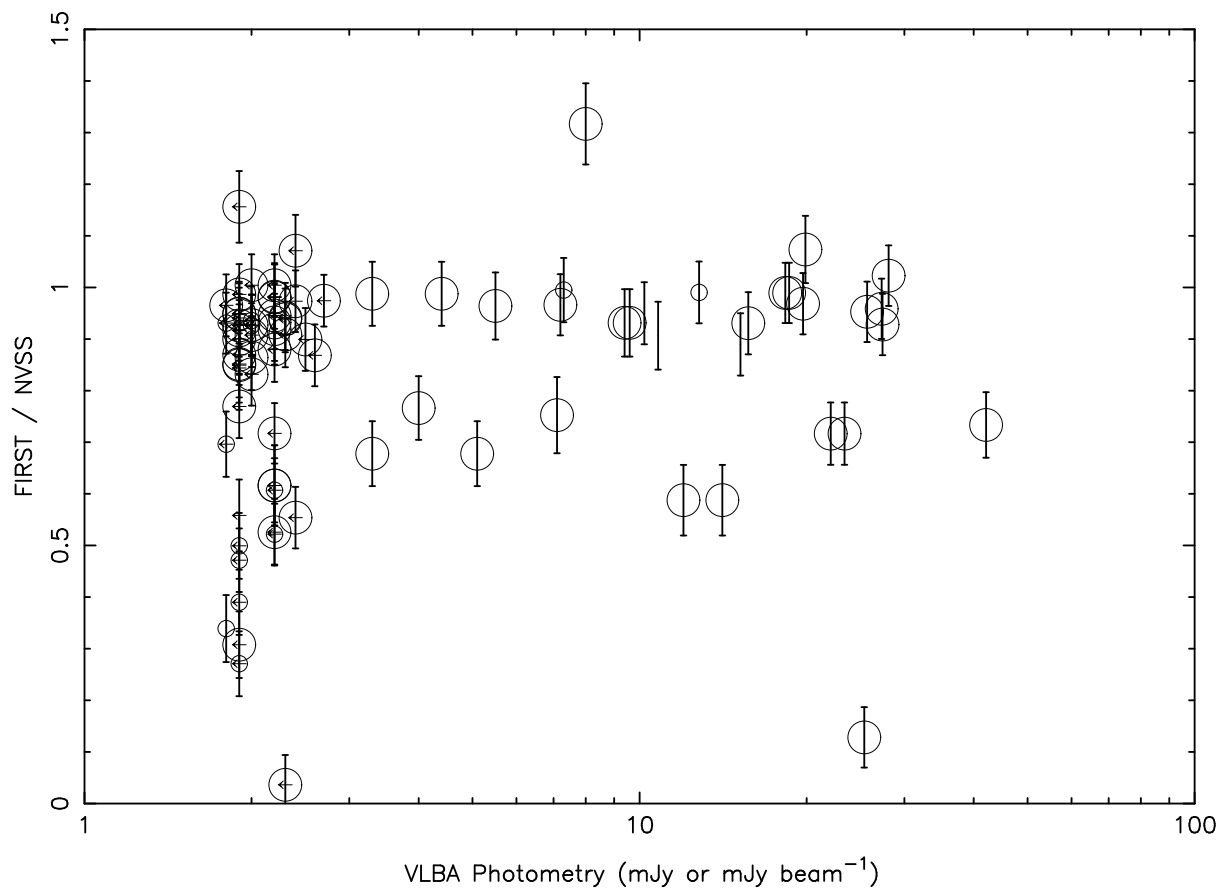}
\caption{Ordinate is the ratio at 1.4~GHz of the peak flux density at
5\arcsec\, resolution from FIRST to the integrated flux density at
45\arcsec\, resolution from the NVSS.  This ratio is a measure of the
compactness of the FIRST source.  Abscissa quantifies the VLBA
photometry at 5.0~GHz, corresponding to the integrated flux density
for VLBA detections, the peak flux density for provisional VLBA
detections, or upper limits to the peak flux density for VLBA
nondetections.  A large circle means the FIRST source is identified in
the optical $I\/$ band, while a small circle means it is
unidentified.}
\end{figure}

\clearpage

\begin{deluxetable}{cccccccccc}
\rotate \tablecolumns{10} \tabletypesize{\scriptsize}
\tablecaption{Faint Sources with Deconvolved Major Diameters Less Than
3\arcsec\, in FIRST\label{tab1}} \tablenum{1} \tablewidth{0pc}
\tablehead{ \multicolumn{3}{c}{VLA at 1.4~GHz }& \colhead{ }&
\multicolumn{5}{c}{VLBA at 5.0~GHz}& \colhead{ }\\ \cline{1-3}
\cline{5-9}\\ \colhead{ }& \colhead{ FIRST }& \colhead{ NVSS }&
\colhead{ }& \colhead{ }& \colhead{ Search }& \colhead{ }& \colhead{
}& \colhead{ }& \colhead{ }\\ \colhead{ FIRST }& \colhead{ Peak
$S_{\rm P}$ }& \colhead{ Integrated $S_{\rm I}$ }& \colhead{ }&
\colhead{ Seg- }& \colhead{ Image $\sigma$ }& \colhead{ R.A.  }&
\colhead{ Decl.  }& \colhead{ Integrated $S_{\rm I}$ }& \colhead{ }\\
\colhead{ Name }& \colhead{ (mJy beam$^{-1}$) }& \colhead{ (mJy) }&
\colhead{ ID\tablenotemark{a} }& \colhead{ ment }& \colhead{ (mJy
beam$^{-1}$) }& \colhead{ (J2000.0) }& \colhead{ (J2000.0) }&
\colhead{ (mJy) }& \colhead{ Note\tablenotemark{b} }\\ \colhead{ (1)}&
\colhead{ (2)}& \colhead{ (3)}& \colhead{ (4)}& \colhead{ (5)}&
\colhead{ (6)}& \colhead{ (7)}& \colhead{ (8)}& \colhead{ (9)}&
\colhead{(10)}\\ } 

\startdata 
 J142456.287$+$352841.80& $ 11.32\pm 0.583$& $ 16.7\pm0.6$& i& 1&
 0.27& 14 24 56.3046& $+$35 28 41.747& $ 3.3\pm0.5$& *\\

 J142456.287$+$352841.80& $ 11.32\pm 0.583$& $ 16.7\pm0.6$& i& 2&
 0.35& 14 24 56.3046& $+$35 28 41.747& $ 5.1\pm0.7$& *\\

 J142516.512$+$345246.62& $ 57.30\pm 2.868$& $ 65.9\pm2.0$& i& 1&
 0.26& ...& ...& $<$1.9\tablenotemark{c}& \\

 J142517.227$+$341559.33& $ 14.34\pm 0.730$& $ 16.3\pm0.6$& i& 2&
 0.31& ...& ...& $<$2.2\tablenotemark{c}& \\

 J142524.214$+$340935.68& $ 15.50\pm 0.790$& $ 20.6\pm1.1$& i& 2&
 0.37& 14 25 24.2201& $+$34 09 35.889& $ 7.1\pm0.7$& *\\

 J142543.885$+$335533.24& $295.30\pm14.766$& $303.1\pm0.9$& i& 2&
 0.38& ...& ...& $<$2.7\tablenotemark{c}& \\

 J142607.716$+$340426.29& $ 33.51\pm 1.682$& $ 34.6\pm1.1$& i& 2&
 0.60& 14 26 07.7162& $+$34 04 26.325& $ 19.7\pm1.2$& *\\

 J142617.948$+$344039.63& $ 11.88\pm 0.611$& $ 38.6\pm1.5$& i& 1&
 0.27& ...& ...& $<$1.9\tablenotemark{c}& *\\

 J142623.374$+$343950.45& $ 16.61\pm 0.842$& $ 42.6\pm1.6$& u& 1&
 0.27& ...& ...& $<$1.9\tablenotemark{c}& *\\

 J142632.164$+$350814.71& $ 87.10\pm 4.357$& $ 95.0\pm2.9$& i& 1&
 0.27& ...& ...& $<$1.9\tablenotemark{c}& \\

 J142659.719$+$341200.21& $ 10.80\pm 0.557$& $297.7\pm7.7$& i& 2&
 0.32& ...& ...& $<$2.3\tablenotemark{c}& *\\

 J142701.644$+$330524.96& $ 16.54\pm 0.834$& $ 18.2\pm0.7$& i& 3a&
 0.32& ...& ...& $<$2.3\tablenotemark{c}& \\

 J142705.073$+$343628.84& $ 27.03\pm 1.358$& $ 32.5\pm1.1$& i& 1&
 0.28& ...& ...& $<$2.0\tablenotemark{c}& \\

 J142738.625$+$330756.96& $ 13.46\pm 0.684$& $ 39.7\pm1.6$& u& 3a&
 0.29& 14 27 38.5713& $+$33 07 56.723&
 $1.8\pm0.3$\tablenotemark{c}&*\\

 J142739.694$+$330744.45& $ 20.87\pm 1.051$& $ 39.7\pm1.6$& i& 3a&
 0.31& ...& ...& $<$2.2\tablenotemark{c}& *\\

 J142744.441$+$333828.62& $ 15.78\pm 0.801$& $ 14.7\pm0.6$& i& 3a&
 0.57& 14 27 44.4413& $+$33 38 28.603& $ 19.9\pm1.2$& *\\

 J142755.936$+$332139.92& $ 15.43\pm 0.783$& $ 17.0\pm0.6$& i& 3a&
 0.28& ...& ...& $<$2.0\tablenotemark{c}& \\

 J142758.722$+$324741.56& $ 17.70\pm 0.893$& $138.0\pm4.1$& i& 3a&
 0.70& 14 27 58.7263& $+$32 47 41.474& $ 25.4\pm1.5$& *\\

 J142806.696$+$325935.83& $ 26.08\pm 1.310$& $ 28.1\pm0.9$& i& 3a&
 0.72& 14 28 06.7043& $+$32 59 35.781& $ 27.4\pm1.7$& *\\

 J142842.556$+$354326.60& $ 19.16\pm 0.968$& $ 19.4\pm0.7$& i& 1&
 0.27& 14 28 42.5556& $+$35 43 26.942& $ 3.3\pm0.6$& *\\

 J142842.556$+$354326.60& $ 19.16\pm 0.968$& $ 19.4\pm0.7$& i& 2&
 0.33& 14 28 42.5556& $+$35 43 26.942& $ 4.4\pm0.8$& *\\

 J142904.545$+$343243.54& $ 12.23\pm 0.628$& $ 14.4\pm1.0$& i& 1&
 0.26& ...& ...& $<$1.9\tablenotemark{c}& *\\

 J142905.105$+$342641.06& $256.22\pm12.812$& $259.0\pm7.8$& i& 2&
 0.42& 14 29 05.1184& $+$34 26 40.997& $ 18.6\pm1.4$& *\\

 J142905.105$+$342641.06& $256.22\pm12.812$& $259.0\pm7.8$& i& 4&
 0.34& 14 29 05.1182& $+$34 26 40.996& $ 18.3\pm1.3$& *\\

 J142906.607$+$354234.67& $ 29.73\pm 1.493$& $ 32.1\pm1.0$& u& 1&
 0.28& ...& ...& $<$2.0\tablenotemark{c}& \\

 J142910.223$+$352946.86& $ 17.49\pm 0.886$& $ 24.4\pm0.8$& i& 1&
 0.50& 14 29 10.2224& $+$35 29 46.891& $ 22.1\pm1.5$& *\\

 J142910.223$+$352946.86& $ 17.49\pm 0.886$& $ 24.4\pm0.8$& i& 2&
 0.63& 14 29 10.2224& $+$35 29 46.891& $ 23.4\pm1.4$& *\\

 J142911.747$+$333144.24& $ 47.34\pm 2.371$& $ 66.0\pm2.0$& i& 3a&
 0.30& ...& ...& $<$2.2\tablenotemark{c}& * \\

 J142937.566$+$344115.69& $ 10.60\pm 0.547$& $ 17.2\pm1.0$& i& 1&
 0.27& 14 29 37.5549& $+$34 41 15.591&
 $2.2\pm0.3$\tablenotemark{c}&*\\

 J142937.566$+$344115.69& $ 10.60\pm 0.547$& $ 17.2\pm1.0$& i& 2&
 0.31& ...& ...& $<$2.2\tablenotemark{c}& *\\

 J143008.885$+$344713.75& $ 26.74\pm 1.345$& $ 27.7\pm0.9$& i& 1&
 0.25& ...& ...& $<$1.8\tablenotemark{c}& \\

 J143018.550$+$341611.65& $ 11.57\pm 0.595$& $ 11.8\pm0.5$& i& 2&
 0.30& ...& ...& $<$2.2\tablenotemark{c}& \\

 J143022.337$+$343727.14& $ 11.28\pm 0.579$& $ 16.2\pm0.6$& u& 1&
 0.25& ...& ...& $<$1.8\tablenotemark{c}& *\\

 J143050.912$+$342614.17& $ 11.46\pm 0.589$& $ 10.7\pm0.5$& i& 2&
 0.34& ...& ...& $<$2.4\tablenotemark{c}& \\

 J143056.092$+$341930.18& $ 33.77\pm 1.694$& $ 64.7\pm2.0$& u& 2&
 0.31& ...& ...& $<$2.2\tablenotemark{c}& *\\

 J143056.256$+$342722.42& $ 12.90\pm 0.659$& $ 13.7\pm0.6$& i& 2&
 0.32& ...& ...& $<$2.3\tablenotemark{c}& \\

 J143109.858$+$335301.67& $ 26.14\pm 1.313$& $ 28.2\pm0.9$& i& 3a&
 0.28& ...& ...& $<$2.0\tablenotemark{c}& *\\

 J143114.097$+$343616.43& $ 16.40\pm 0.834$& $ 16.7\pm0.6$& i& 2&
 0.31& ...& ...& $<$2.2\tablenotemark{c}& \\

 J143121.320$+$332808.95& $ 27.33\pm 1.373$& $ 27.6\pm0.9$& u& 3a&
 0.35& 14 31 21.3546& $+$33 28 08.744& $ 12.8\pm1.2$& *\\

 J143123.297$+$331625.82& $ 15.38\pm 0.781$& $ 16.5\pm0.9$& i& 3a&
 0.31& 14 31 23.3356& $+$33 16 25.880&
 $2.2\pm0.3$\tablenotemark{c}&*\\

 J143134.549$+$351511.19& $ 58.47\pm 2.927$& $ 76.0\pm2.7$& i& 1&
 0.27& ...& ...& $<$1.9\tablenotemark{c}& *\\

 J143152.544$+$340110.15& $ 13.69\pm 0.698$& $ 14.7\pm0.6$& i& 2&
 0.36& 14 31 52.5521& $+$34 01 09.990& $ 9.6\pm1.1$& *\\

 J143152.544$+$340110.15& $ 13.69\pm 0.698$& $ 14.7\pm0.6$& i& 4&
 0.30& 14 31 52.5520& $+$34 01 09.990& $ 9.4\pm0.9$& *\\

 J143155.856$+$345926.99& $ 10.53\pm 0.545$& $ 11.7\pm0.5$& i& 1&
 0.27& ...& ...& $<$1.9\tablenotemark{c}& \\

 J143213.543$+$350941.05& $ 15.65\pm 0.796$& $ 16.5\pm0.6$& i& 1&
 0.27& ...& ...& $<$1.9\tablenotemark{c}& \\

 J143222.785$+$324939.62& $ 11.63\pm 0.592$& $ 12.3\pm0.5$& i& 3a&
 0.31& ...& ...& $<$2.2\tablenotemark{c}& \\

 J143239.561$+$350151.42& $ 12.86\pm 0.659$& $ 15.1\pm0.6$& i& 1&
 0.26& ...& ...& $<$1.9\tablenotemark{c}& \\

 J143256.072$+$353339.52& $ 62.94\pm 3.150$& $ 66.3\pm2.0$& i& 1&
 0.27& ...& ...& $<$1.9\tablenotemark{c}& \\

 J143309.671$+$351520.14& $ 13.77\pm 0.704$& $ 29.2\pm1.0$& u& 1&
 0.26& ...& ...& $<$1.9\tablenotemark{c}& *\\

 J143311.054$+$335828.81& $ 19.54\pm 0.987$& $ 25.5\pm0.9$& i& 4&
 0.25& 14 33 11.0825& $+$33 58 28.533& $ 4.0\pm0.5$& *\\

 J143345.947$+$341149.03& $ 16.82\pm 0.852$& $ 62.1\pm2.3$& u& 4&
 0.26& ...& ...& $<$1.9\tablenotemark{c}& *\\

 J143428.007$+$331102.05& $ 21.86\pm 1.103$& $ 23.6\pm0.8$& i& 4&
 0.26& ...& ...& $<$1.9\tablenotemark{c}& \\

 J143432.836$+$352141.47& $ 20.32\pm 1.024$& $ 40.7\pm1.6$& u& 1&
 0.26& ...& ...& $<$1.9\tablenotemark{c}& *\\

 J143434.217$+$351009.53& $ 66.72\pm 3.339$& $ 69.6\pm2.1$& i& 2&
 0.77& 14 34 34.2216& $+$35 10 09.432& $ 27.3\pm1.9$& *\\

 J143435.361$+$352622.26& $ 30.93\pm 1.552$& $ 30.8\pm1.0$& i& 1&
 0.28& ...& ...& $<$2.0\tablenotemark{c}& \\

 J143435.361$+$352622.26& $ 30.93\pm 1.552$& $ 30.8\pm1.0$& i& 2&
 0.30& ...& ...& $<$2.2\tablenotemark{c}& \\

 J143445.321$+$332820.58& $ 26.48\pm 1.331$& $ 36.1\pm1.4$& i& 4&
 0.97& 14 34 45.3500& $+$33 28 20.571& $ 42.1\pm2.2$& *\\

 J143446.491$+$332452.27& $ 12.04\pm 0.616$& $ 13.1\pm0.6$& i& 4&
 0.27& 14 34 46.5226& $+$33 24 52.151&
 $2.2\pm0.3$\tablenotemark{c}&*\\

 J143449.111$+$354246.98& $ 21.57\pm 1.088$& $ 36.7\pm1.7$& i& 1&
 0.32& 14 34 49.1016& $+$35 42 47.221& $ 12.0\pm1.1$& *\\

 J143449.111$+$354246.98& $ 21.57\pm 1.088$& $ 36.7\pm1.7$& i& 2&
 0.38& 14 34 49.1016& $+$35 42 47.221& $ 14.1\pm1.3$& *\\

 J143453.497$+$343626.65& $ 24.43\pm 1.230$& $ 25.1\pm0.8$& i& 2&
 0.34& ...& ...& $<$2.4\tablenotemark{c}& \\

 J143507.129$+$331933.55& $ 13.38\pm 0.683$& $ 14.3\pm0.6$& u& 4&
 0.28& ...& ...& $<$2.0\tablenotemark{c}& \\

 J143525.184$+$325124.06& $ 11.56\pm 0.587$& $ 13.7\pm0.6$& ...& 4&
 0.26& ...& ...& $<$1.9\tablenotemark{c}& \\

 J143527.951$+$331145.46& $ 38.28\pm 1.918$& $ 44.3\pm1.7$& i& 4&
 0.26& 14 35 27.9710& $+$33 11 45.456&
 $2.0\pm0.3$\tablenotemark{c}&*\\

 J143528.374$+$331931.53& $ 22.52\pm 1.135$& $ 24.2\pm0.8$& i& 4&
 0.41& 14 35 28.3946& $+$33 19 31.464& $ 15.7\pm1.1$& *\\

 J143529.161$+$343422.94& $ 42.71\pm 2.140$& $ 44.2\pm1.4$& i& 4&
 0.27& 14 35 29.1470& $+$34 34 22.721& $ 7.2\pm0.9$& *\\

 J143539.740$+$344359.55& $ 21.45\pm 1.081$& $ 24.7\pm0.8$& i& 2&
 0.36& ...& ...& $<$2.6\tablenotemark{c}& \\

 J143543.629$+$353305.02& $ 17.56\pm 0.888$& $ 31.7\pm1.0$& i& 2&
 0.34& ...& ...& $<$2.4\tablenotemark{c}& *\\

 J143643.209$+$352222.98& $ 19.10\pm 0.965$& $ 19.2\pm0.7$& u& 2&
 0.38& 14 36 43.2194& $+$35 22 22.963& $ 7.3\pm1.0$& *\\

 J143647.710$+$331037.66& $ 11.89\pm 0.609$& $ 21.3\pm1.0$& ...& 4&
 0.26& ...& ...& $<$1.9\tablenotemark{c}& *\\

 J143713.571$+$350554.84& $ 56.78\pm 2.843$& $ 59.6\pm1.8$& i& 3b&
 0.66& 14 37 13.5686& $+$35 05 54.832& $ 25.7\pm1.8$& *\\

 J143723.715$+$350734.69& $ 98.55\pm 4.930$& $109.6\pm3.8$& i& 3b&
 0.35& ...& ...& $<$2.5\tablenotemark{c}& *\\

 J143728.413$+$331110.20& $ 11.15\pm 0.577$& $ 12.3\pm0.5$& ...& 4&
 0.35& 14 37 28.4307& $+$33 11 10.069& $ 10.8\pm0.8$& *\\

 J143737.176$+$342006.52& $ 38.91\pm 1.950$& $ 41.8\pm1.3$& u& 4&
 0.25& ...& ...& $<$1.8\tablenotemark{c}& \\

 J143739.458$+$325408.08& $ 40.08\pm 2.007$& $ 42.6\pm1.3$& ...& 4&
 0.27& ...& ...& $<$1.9\tablenotemark{c}& \\

 J143752.050$+$351940.08& $ 11.06\pm 0.568$& $ 8.4\pm0.5$& i& 3b&
 0.34& 14 37 52.0516& $+$35 19 39.965& $ 8.0\pm0.9$& *\\

 J143811.062$+$340459.63& $136.04\pm 6.804$& $137.9\pm4.2$& i& 4&
 0.27& ...& ...& $<$1.9\tablenotemark{c}& \\

 J143821.829$+$344001.06& $ 12.37\pm 0.634$& $ 10.7\pm0.5$& i& 4&
 0.26& ...& ...& $<$1.9\tablenotemark{c}& \\

 J143841.949$+$335809.48& $ 57.17\pm 2.862$& $ 55.9\pm1.7$& i& 4&
 0.48& 14 38 41.9460& $+$33 58 09.504& $ 28.1\pm1.7$& *\\

 J143850.267$+$340419.84& $ 12.05\pm 0.618$& $ 12.5\pm0.5$& i& 4&
 0.30& 14 38 50.2712& $+$34 04 20.173& $ 5.5\pm0.6$& *\\

 J143850.280$+$350021.16& $ 43.82\pm 2.195$& $ 46.7\pm1.5$& i& 3b&
 0.32& ...& ...& $<$2.3\tablenotemark{c}& \\

 J143859.469$+$345309.33& $ 13.84\pm 0.709$& $ 22.8\pm0.8$& u& 3b&
 0.31& ...& ...& $<$2.2\tablenotemark{c}& *\\

 J143909.406$+$332101.73& $ 84.25\pm 4.215$& $ 94.7\pm3.2$& ...& 4&
 0.35& 14 39 09.4176& $+$33 21 01.674& $ 15.2\pm1.4$& *\\

 J143916.249$+$344912.09& $ 31.63\pm 1.589$& $ 33.3\pm1.1$& ...& 3b&
 0.39& 14 39 16.2428& $+$34 49 11.970& $ 10.2\pm0.9$& *\\
\enddata 

\tablecomments{Units of right ascension are hours, minutes, and
 seconds, and units of declination are degrees, arcminutes, and
 arcseconds. One-dimensional errors at 1 $\sigma$ are estimated to be
 2.5~mas.}
\tablenotetext{a}{Identification status in the $I\/$ band from Jannuzi
 et al.\ 2005.  i means identified with a galaxy or quasar.  u means
 unidentified.}
\tablenotetext{b}{An * means there is a note on the source in the
 appendix.}
\tablenotetext{c}{Peak flux density at 2 mas resolution after
 correction for coherence losses.  Upper limits are 6 $\sigma$.}
\end{deluxetable}

\clearpage

\begin{deluxetable}{cccclcccccc}
\rotate
\tablecolumns{11}
\tabletypesize{\scriptsize} 
\tablecaption{Calibrators in Frame of IRCF-Ext.1 Catalog\label{tab2}}
\tablenum{2}
\tablewidth{0pc}
\tablehead{ 
\colhead{            }&
\colhead{            }&
\colhead{            }&
\multicolumn{5}{c}{Calibrator}&
\colhead{            }&
\colhead{            }&
\colhead{            }\\
\cline{4-8}\\
\colhead{ VLBA       }&
\colhead{            }&
\colhead{ Number     }&
\colhead{            }&
\colhead{            }&
\colhead{            }&
\colhead{            }&
\colhead{            }&
\colhead{ Switching  }&
\colhead{ Coherence  }&
\colhead{ Position   }\\
\colhead{ Seg-       }&
\colhead{            }&
\colhead{ of         }&
\colhead{            }&
\colhead{            }&
\colhead{ R.A.       }&
\colhead{ Decl.      }&
\colhead{            }&
\colhead{ Angle      }&
\colhead{ Correc-    }&
\colhead{ Correc-    }\\
\colhead{ ment       }&
\colhead{ UT Epoch   }&
\colhead{ Antennas   }&
\colhead{ Type       }&
\colhead{ Name       }&
\colhead{ (J2000.0)  }&
\colhead{ (J2000.0)  }&
\colhead{ Ref.       }&
\colhead{ (\arcdeg)  }&
\colhead{ tion       }&
\colhead{ tion (mas) }\\
\colhead{ (1)}& \colhead{ (2)}& \colhead{ (3)}& \colhead{ (4)}& 
\colhead{ (5)}& \colhead{ (6)}& \colhead{ (7)}& \colhead{ (8)}& 
\colhead{ (9)}& \colhead{(10)}& \colhead{(11)}\\
}
\startdata
1 & 2001 Apr 26 & 10 & 
      Phase& J1426$+$3625& 14 26 37.08749& $+$36 25 09.5739& 1  & ...& ...& ...\\
& & & Check& J1416$+$3444& 14 16 04.18624& $+$34 44 36.4276& 1,2& 2.7& 1.2& $-$1.2,$-$0.2\\
& & & Check& F14543631   & 14 32 39.82965& $+$36 18 07.9321& 1  & 1.2& 1.1& $+$0.2,$-$1.5\\

2 & 2002 Apr 26 &  9 & 
      Phase& J1426$+$3625& 14 26 37.08749& $+$36 25 09.5739& 1  & ...& ...& ...\\
& & & Check& J1416$+$3444& 14 16 04.18624& $+$34 44 36.4276& 1,2& 2.7& 1.4& $-$0.1,$-$0.3\\
& & & Check& J1438$+$3710& 14 38 53.61097& $+$37 10 35.4168& 1,2& 2.6& 1.2& $-$0.5,$+$0.1\\
& & & Check& F14543631   & 14 32 39.82965& $+$36 18 07.9321& 1  & 1.2& 1.1& $+$0.1,$-$1.4\\

3a& 2002 Apr 28 &  9 & 
      Phase& J1422$+$3223& 14 22 30.37894& $+$32 23 10.4405& 1  & ...& ...& ...\\
& & & Check& J1416$+$3444& 14 16 04.18624& $+$34 44 36.4276& 1,2& 2.7& 1.1& $+$0.4,$-$1.0\\

3b& 2002 Apr 28 &  9 & 
      Phase& J1438$+$3710& 14 38 53.61097& $+$37 10 35.4168& 1,2& ...& ...& ...\\
& & & Check& J1426$+$3625& 14 26 37.08749& $+$36 25 09.5739& 1  & 2.6& 1.2& $+$0.2,$+$0.3\\
& & & Check& F14543631   & 14 32 39.82965& $+$36 18 07.9321& 1  & 1.5& 1.1& $-$0.6,$-$1.2\\

4 & 2002 Jun 15 \tablenotemark{a} & 10 & 
      Phase& J1442$+$3234& 14 42 00.13897& $+$32 34 20.3016& 1,2& ...& ...& ...\\
& & & Check& F14653289   & 14 39 23.65450& $+$32 53 54.8237& 1  & 0.6& 1.0& $+$2.3,$-$0.9\\
\enddata
\tablecomments{Units of right ascension are hours, minutes, and
seconds, and units of declination are degrees, arcminutes, and
arcseconds.  One-dimensional errors at 1 $\sigma$ are better than
1~mas for all calibrators except F14653289, for which they are better
than 1.3~mas.}
\tablenotetext{a}{FIRST J143445.321$+$332820.58 was also strong enough
to phase self-calibrate and thus serve as a coherence check.  It was
observed with a switching angle of 1.8\arcdeg\, and yielded a
coherence correction of 1.1.}
\tablerefs{(1) N. Vandenberg 2001, private commmunication; (2) Beasley
et al.\ 2002.}
\end{deluxetable}
\end{document}